\documentstyle[prl,aps,multicol]{revtex}
\input{epsf}
\begin{document}
\draft
\title{Quantum opening of the Coulomb gap in two dimensions} 
\author{Giuliano Benenti$^{(a)}$, Xavier Waintal$^{(a)}$, 
Jean-Louis Pichard$^{(a)}$, and Dima L. Shepelyansky$^{(b)}$} 
\address{$^{(a)}$CEA, Service de Physique de l'Etat Condens\'e, 
Centre d'Etudes de Saclay, F-91191 Gif-sur-Yvette, France}  
\address{$^{(b)}$Laboratoire de Physique Quantique, UMR C5626
du CNRS, Universit\'e Paul Sabatier, 31062 Toulouse, France} 
\date{\today}
\maketitle

\begin{abstract}
 For a constant density of spinless fermions in strongly 
disordered two dimensional clusters, the energy level spacing between 
the ground state and the first excitation is studied 
for increasing system sizes. The average indicates a smooth opening of 
the gap when the Coulomb energy to Fermi energy ratio $r_s$ increases 
from $0$ to $3$, while the distribution exhibits a sharp Poisson-Wigner-like  
transition at $r_s \approx 1$. The results are related
to the transition from Mott to Efros-Shklovskii hopping conductivity 
recently observed at a similar ratio $r_s$.

\end{abstract}

\begin{multicols}{2}
\narrowtext

 In disordered insulators, a crossover \cite{ES} in the temperature 
dependence of the resistivity $\rho(T)$ is induced by Coulomb interactions
from the Mott variable range hopping law ($\rho(T)=\rho_M \exp (T_0/T)^{1/3}$ 
in dimension $d=2$) to the Efros-Shklovskii behavior 
($\rho(T) = \rho_{\rm{ES}} \exp (T_{\rm{ES}}/T)^{1/2}$). The long range 
nature of the interactions leads to a dip in the single particle density 
of states, and the assumption that single electron hopping dominates the 
transport leads to this change in the resistivity. However, a single electron 
hop may reorganize the location of the other particles, inducing complex 
many particle excitations. This makes the Coulomb gap problem 
difficult and gives us the motivation to study the first quantum excitation 
above the ground state. For $d=2$, the strength of Coulomb interactions is 
very often given in units of the Fermi energy by the dimensionless ratio 
$r_s$. For strong disorder, the first excitation energy is expected to 
become larger when $r_s$ increases, and to decay as $1/L$ (instead of 
$1/L^2$ for free electrons) when the system size $L$ increases. But 
one cannot estimate the threshold $r_s^C$ where the Coulomb gap 
opens without taking into account all complex of many body quantum 
processes. Considering spinless fermions in $2d$ strongly 
disordered clusters, we confirm from numerical calculations that 
the gap opens at a value $r_s^C \approx 1.2$, and we point out that this 
is indeed for a similar value ($r_s\approx 1.7$) that a change in the hopping 
conductivity from Mott hopping to Coulomb gap behavior has been 
reported \cite{khondaker} for an electron gas created at a 
GaAs/AlGaAs heterostructure. For a statistical ensemble of clusters, we 
have calculated the many body states at the mean field level given by 
the Hartree-Fock approximation and we have added the effects of the 
residual interaction. Keeping constant the carrier density $n_e$, and 
increasing the size $L$, the average gap between the ground state 
and the first excitation behaves as $1/L^{\alpha}$, with $\alpha$ decreasing 
from $2$ to $1$ when $r_s$ increases from $0$ to $3$. Another remarkable 
effect of the interaction is to yield a sharp transition for the 
gap distribution: it tends to Poisson or to Wigner-like distributions 
for small or large $r_s$ respectively at the thermodynamic limit. 
A critical threshold $r_s^C \approx 1.2$ is characterized by a scale 
invariant gap distribution, reminiscent of the one particle problem 
\cite{Shklovskii} at a mobility edge. However, it is only the 
distribution of the first spacing which exhibits such a transition, the 
distributions of the next spacings remain Poissonian and
are essentially unchanged when $r_s$ varies. 
Eventually, we discuss the implications for the 
hopping conductivity and we confirm that the transition for the gap takes 
place at a smaller $r_s$ than $r_s^F \approx 4-5$ where a change in 
the topology of the persistent currents carried by the ground state 
has been observed \cite{paper1,note}. 

 We consider a disordered square lattice with $M=L^2$ sites occupied by 
$N$ spinless fermions. The Hamiltonian reads 
\begin{equation} 
\label{hamiltonian} 
H=-t\sum_{<i,j>} c^{\dagger}_i c_j +  
\sum_i v_i n_i  + U \sum_{i\neq j} \frac{n_i n_j } {2r_{ij}},
\end{equation} 
where $c^{\dagger}_i$ ($c_i$) creates (destroys) an electron in 
the site $i$, the hopping term $t$ between nearest neighbours 
characterizes the kinetic energy, $v_i$ the site potentials 
taken at random inside the interval $[-W/2,+W/2]$, 
$n_i=c^{\dagger}_i c_i$ is the occupation number at site $i$ and $U$ 
measures the strength of the Coulomb repulsion. The boundary conditions 
are periodic and $r_{ij}$ is the inter-particle distance for a $2d$ torus. 
If $a^*_B=\hbar^2\epsilon/(m^* e^2)$, $m^*$, $\epsilon$, $a$ and 
$n_s=N/(a L^2)$ denote respectively the effective Bohr radius, 
the effective mass, the dielectric constant, the lattice spacing  
and the carrier density, the factor $r_s$ is given by:
\begin{equation}
r_s=\frac{1}{\sqrt{\pi n_s} a^*_B} = \frac{U}{2t\sqrt{\pi n_e}}, 
\end{equation} 
since in our units $\hbar^2/(2m^*a^2)\to t$, $e^2/(\epsilon a)\to U$ 
and $n_e=N/L^2$. 

 In this study, a large disorder to hopping ratio $W/t=15$ 
is imposed for having Anderson localization and Poissonian spectral 
statistics for the one particle levels at $r_s=0$ when $L \geq 8$. 
We study $N=4,9$ and $16$ particles inside clusters of size $L=8,12$ 
and $16$ respectively. This corresponds to a constant low carrier 
density $n_e=1/16$. 
A numerical study via exact diagonalization techniques for sparse matrices 
is possible only for small systems \cite{paper1}, and does not allow us 
to vary $L$ for a constant density. We are obliged to look for an 
approximate solution of the problem, using the Hartree-Fock (HF) orbitals, 
and to control the validity of the approximations. One starts from the 
HF Hamiltonian where the two-body part is reduced to an effective single 
particle Hamiltonian 
\cite{Kato,Poilblanc,Schreiber}
\begin{eqnarray} 
\label{hfhamiltonian} 
\begin{array}{c}
\displaystyle{ 
U (\sum_{i\neq j} \frac{1}{r_{ij}} n_i\langle n_j \rangle
- \sum_{i\neq j} \frac{1}{r_{ij}} c^{\dagger}_i c_j  
\langle c^{\dagger}_j c_i \rangle }),  
\label{HF}
\end{array} 
\end{eqnarray} 
where $\langle ... \rangle$ stands for the expectation value with 
respect to the HF ground state, which has to be determined 
self-consistently. For large values of the interaction 
and large system sizes the single-particle problem (\ref{hfhamiltonian}) 
is still non-trivial, since the self-consistent iteration can be trapped 
in metastable states. This limits our study to small $r_s$ 
and forbids us to study by this method charge crystallization 
discussed in \cite{paper1} at a larger $r_s^W \approx 12$ .      

 The mean field HF results can be improved using a method 
\cite{CIphys1,CIphys2} known as the configuration interaction method (CIM) 
in quantum chemistry \cite{CIchem}. Once a complete orthonormal basis of 
HF orbitals has been calculated 
($H_{HF} |\psi_{\alpha}\rangle = \epsilon_{\alpha}|\psi_{\alpha}\rangle$   
with $ \alpha=1,2,\ldots,L^2$), it is possible to build up a Slater 
determinants' basis for the many-body problem which can be truncated to 
the $N_H$ first Slater determinants, ordered by increasing energies. 
The two-body Hamiltonian can be written as 
\begin{eqnarray}
H_{\rm int}=\frac{1}{2} \sum_{\alpha,\beta,\gamma,\delta}  
Q_{\alpha\beta}^{\gamma\delta} d_{\alpha}^{\dagger} d_{\beta}^{\dagger} 
d_{\delta}d_{\gamma} ,
\label{Hint}
\end{eqnarray} 
with 
\begin{eqnarray} 
Q_{\alpha\beta}^{\gamma\delta} = 
U\sum_{i\neq j} \frac{ \psi_\alpha(i) \psi_\beta(j) \psi_\gamma(i) 
\psi_\delta (j)}{r_{ij}}  
\end{eqnarray}
and $d^{\dagger}_{\alpha}=\sum_j \psi_{\alpha} (j) c^{\dagger}_j |0 \rangle$. 
One gets the residual interaction subtracting Eq. \ref{HF} from  
Eq. \ref{Hint}. This keeps the two-body nature of 
the Coulomb interaction, and if $N\gg 2$ it is still possible to take 
advantage of the sparsity of the matrix and to diagonalize it via the 
Lanczos algorithm.  

\begin{figure}
\centerline{
\epsfxsize=8cm 
\epsfysize=7.4cm 
\epsffile{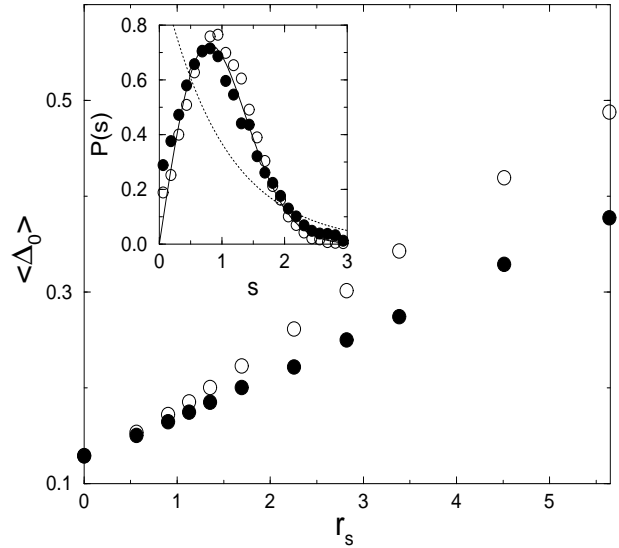}
}
\caption{ 
CIM result (full circle) compared to HF approximation (empty circle)
for $N=9$ and $L=12$. Mean gap $< \Delta_0 >$ and gap distribution 
at $r_s=4.5$ (insert).
}
\label{fig1} 
\end{figure}

  We have first compared HF and CIM results. Labelling the levels 
by increasing energy and studying an ensemble of $10^4$ samples, 
we have studied the first spacing $\Delta_0=E_1-E_0$. The role of 
the residual interaction can be seen in Fig. \ref{fig1}. When 
$r_s > 1$, the residual interaction reduces the mean gap, and 
slightly changes the distribution. The CIM results agree with the 
results given from exact diagonalization with an accuracy of the 
order $2\%$ when one takes into account the 
$N_H=10^3$ first Slater determinants when $r_s=5$ and $L=8$. This 
means that a basis spanning only $0.2\%$ of the total Hilbert space 
is sufficient for studying the first excitations. For larger $L$, 
exact diagonalization is no longer possible, but one can look if 
the results vary when $N_H$ increases. 
In the worst case considered ($L=16$, $r_s=2.8$) the accuracy in the first 
four spacings can be estimated of the order $5\%$ when 
$N_H=2\times 10^3$. 

Therefore the CIM method allows to study low energy level statistics
for $r_s<3$. 
However, its accuracy is not sufficient to determine 
a small change of the ground state energy 
when the boundary conditions are twisted (i.e. the persistent currents).

\begin{figure} 
\centerline{
\epsfxsize=8cm 
\epsfysize=7.4cm 
\epsffile{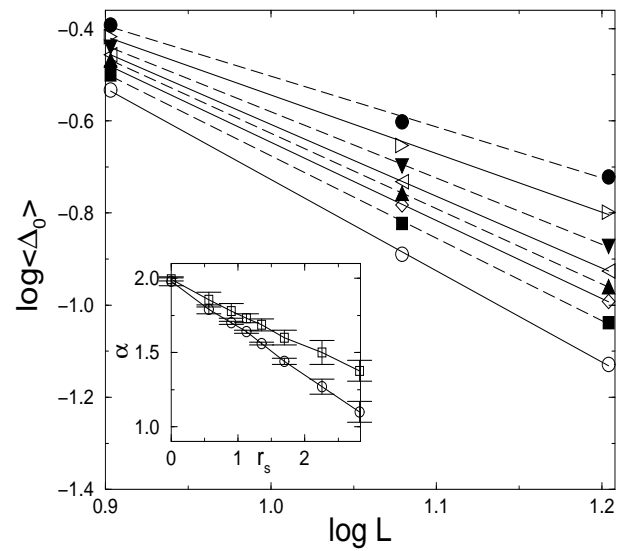}
}
\caption{ 
Size dependence of the average gap 
(first spacing $<\Delta_0> \propto L^{-\alpha (r_s)}$). 
From bottom to top: $r_s=0,0.6,0.9,1.1,1.4,1.7,2.3,2.8$. 
Insert: $\alpha (r_s)$ (circle, characterizing $<\Delta_0>$ and 
square, average over $i=1-3$, characterizing $<\Delta_i>$ ).    
}
\label{fig2} 
\end{figure} 
 
 We have calculated the first energy levels for different 
sizes $L$. The first average spacing $<\Delta_0>$ calculated for an ensemble 
of $10^4$ samples is given in Fig. \ref{fig2}. It exhibits a power law 
decay as $L$ increases, with an exponent 
$\alpha$ given in the insert. One finds for the first spacing 
that $\alpha$ linearly decreases from $d=2$ to $1$ when $r_s$ 
increases from $0$ to $3$. This proves a gradual opening of the mean 
Coulomb gap. The next mean spacings depend more weakly on 
$r_s$, as shown in Fig. \ref{fig2}. 

\begin{figure} 
\centerline{
\epsfxsize=8cm 
\epsfysize=7.4cm 
\epsffile{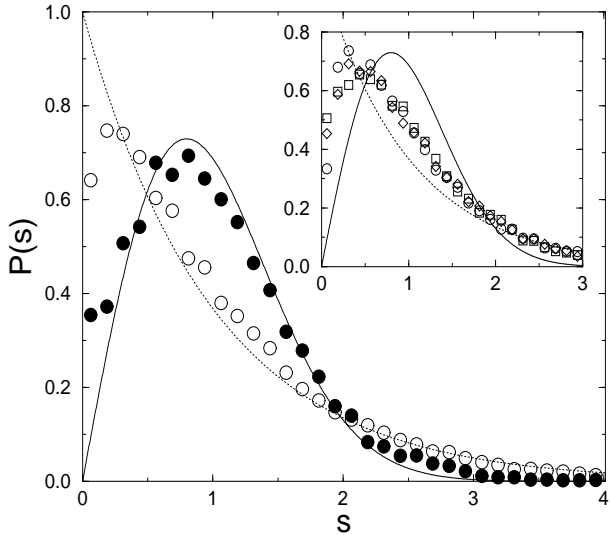}
}
\caption{Gap distribution $P(s)$ for $r_s=0$ (empty circle) and 
$r_s=2.8$ (full circle) when $L=16$, compared to $P_P(s)$ and 
$P_W(s)$. Insert: size invariant $P(s)$ at $r_s^C\approx 1.2$ ; 
$L=8$ (circle), $12$ (square) and $16$ (diamond).
}
\label{fig3} 
\end{figure} 

\begin{figure} 
\centerline{
\epsfxsize=8cm 
\epsfysize=7.4cm 
\epsffile{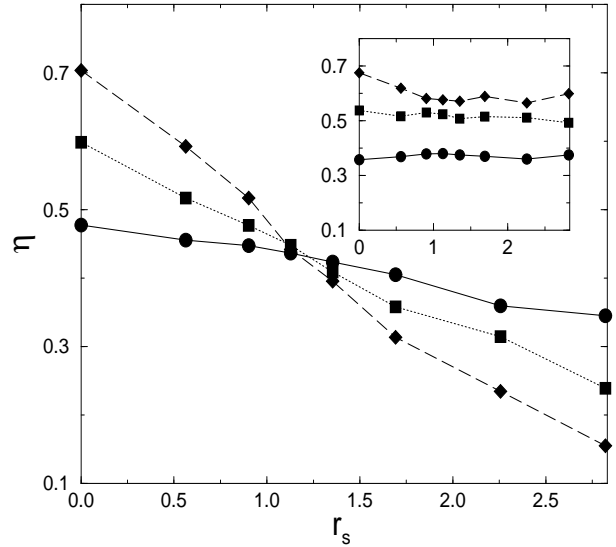}
}
\caption{ 
Parameter $\eta(r_s)$ corresponding to the first spacing $\Delta_0$ 
at $L=8$ (circle), $12$ (square) and $16$ (diamond). Insert: $\eta(r_s)$ 
for the second spacing $\Delta_1$. 
}
\label{fig4} 
\end{figure}

For $r_s=0$, the distribution of the first spacing $s=\Delta_0/<\Delta_0>$
becomes more and more close to the Poisson distribution $P_P(s)=\exp(-s)$ 
when $L$ increases, as it should be for an Anderson insulator. For a 
larger $r_s$, the distribution seems to become close to the Wigner surmise 
$P_W(s)=(\pi s /2) \exp( -\pi s^2 /4)$ characteristic of level repulsion 
in random matrix theory, as  shown for $r_s = 2.8$ and $L=16$ for instance. 
To study how this $P(s)$ goes from 
Poisson to a Wigner-like distribution when $r_s$ increases, we have 
calculated a parameter $\eta$ which decreases from $1$ to $0$ when 
$P(s)$ goes from Poisson to Wigner:
\begin{eqnarray} 
\eta=\frac{\hbox{var}(P(s))-\hbox{var}(P_W(s))} 
{\hbox{var}(P_P(s))-\hbox{var}(P_W(s))}, 
\end{eqnarray}  
where $\hbox{var}(P(s))$ denotes the variance of $P(s)$, 
$\hbox{var}(P_P(s))=1$ and $\hbox{var}(P_W(s))=0.273$.
In Fig.\ref{fig4}, one can see that three curves $\eta (r_s)$ 
characterizing the first spacing for $L=8, 12, 16$ intersect at 
a critical value $r_s^C \approx 1.2$. For $r_s < r_s^C$ the 
distribution tends to Poisson in the thermodynamic limit, while 
for $r_s>r_s^C$ it tends to a Wigner-like behavior 
\cite{glass}. At the threshold 
$r_s^C$, there is a size-independent intermediate distribution shown 
in the insert of Fig. \ref{fig3}, exhibiting level repulsion 
at small $s$ followed by a $\exp( -a s)$ decay at large $s$ 
with $a \approx 1.52$. This Poisson-Wigner transition characterizes 
only the first spacing, the distributions of the next spacings being 
quite different. The insert of Fig. \ref{fig4} does not show an intersection 
for the parameter $\eta$ calculated with the second spacing. 

 The second excitation is less localized than the first one when $r_s=0$, 
since the one particle localization length weakly increases with 
energy. This is only for $L=16$ that the distribution of the 
second spacing becomes close to Poisson without interaction, and a weak  
level repulsion occurs as $r_s$ increases. The observed transition, and 
the difference between the first spacing and the following ones is mainly 
an effect of the HF mean field. For the first spacing, the curves $\eta$ 
calculated with the HF data are qualitatively the same. 
At the mean field level the first excitation is a particle-hole 
excitation starting from the ground state and requires an energy of the 
order $U/L$, with fluctuactions around this mean value. The second excitation 
is again a particle-hole excitation starting from the ground state. The 
energy spacing between the first and the second excited state is given by the 
difference of two uncorrelated particle-hole excitations and a
Poissonian distribution follows naturally. 
For $r_s > r_s^C$, the Gaussian-like HF distributions for 
$\Delta_0$ become more Wigner-like when the residual interaction is included.
We point out that in metallic quantum dots also, the first excitation is 
statistically different from the others, as shown by numerical studies 
\cite{walker} within the HF approximation. 

 The existence of a critical $r_s$ value for the opening of the Coulomb
gap can be understood similarly to \cite{PD99}. The single particle 
density of states around the Fermi energy $E_F$ is given by 
\cite{ES} $\rho(E)\approx |E-E_F|/U^2$ and the gap size 
$\Delta_g=|E_g-E_F|$ can be estimated from the condition 
$\rho(E_g)\approx\overline{\rho}$, 
with $\overline{\rho}\approx 1/W$ mean density of states for $W\gg t$, 
obtaining $\Delta_g\approx U^2/W$.  
According to Fermi golden rule, the inverse lifetime of a Slater 
determinant built from electrons localized at given sites is   
$\Gamma_t\approx t^2 (1/W) (N/L^2)$, with $N/(W L^2)$ density 
of states directly coupled by the hopping term of the Hamiltonian 
(\ref{hamiltonian}). 
Therefore at zero temperature quantum fluctuactions melt the Coulomb 
gap for $\Gamma_t\approx\Delta_g$, giving $r_s\approx r_s^C\approx 1$. 
We conclude that a crossover from Efros-Shklovskii to Mott hopping 
conductivity is expected not only increasing temperature but also 
increasing carrier density, as observed in \cite{khondaker}.   

 To measure possible delocalization effects, we have calculated 
the number of occupied sites per particle $\xi_s=N/\sum_{i} \rho^{2}_i$ 
where $\rho_i=\langle\Psi_0|n_i|\Psi_0\rangle$ is the charge density of 
the ground state at the site $i$.  
Around $r_s \approx 1.2$ and after 
ensemble average, the maximum increase of $\xi_s$  compared to $r_s=0$
is negligibly small ($2 \%$). These are mainly the distribution 
and the average value of the first excitation energy which exhibit noticeable 
effects. This 
matters for the hopping conductivity. The usual argument is to consider 
the length $L(T)$ where $\exp[ -({2L/\xi(r_s)}+\Delta_0(r_s)/kT)]$ is 
maximum with the localization length $\xi (r_s) \approx \xi(0)$.
If one takes for $\Delta_0(r_s)$ its average value $\approx (A+B r_s)
/L^{\alpha(r_s)}$ (see Fig. \ref{fig1}), one obtains for the hopping 
resistivity a smooth and continuous crossover from Mott to Efros-Shklovskii 
hopping, given by: 
\begin{equation}
\rho(T)\propto \exp \left(\frac{T(r_s)}{T}\right)^{1/(\alpha+1)}, 
\end{equation} 
where 
\begin{equation}
T(r_s) \approx \frac{A+B r_s}{k \xi^{\alpha}}.
\end{equation}
This prediction neglects the sharp transition in the distribution 
of $\Delta_0$ at $r_s^C$, which could be better included by considering 
a more typical value for $\Delta_0 (r_s)$ than its average, for instance 
obtained from the value $s_b$ for which $\int_0^{s_b} p(s) ds = b$, with 
$b=1/2$ for instance. This will introduce 
a sharp discontinuity at $r_s^C$ in $T(r_s)$. 
 
In summary, we have analyzed the Coulomb gap statistics for spinless 
fermions in a strongly disordered squared lattice when $r_s < 3$. 
On one hand, we have found a sharp interaction-induced transition 
at $r_s^C\approx 1.2$, characterized by a scale invariant distribution. 
Around the critical point, the gap distribution tends to Poisson or 
to Wigner-like distributions respectively at the thermodynamic limit. 
This effect is present at the HF mean field level, the residual 
interaction weakly shifts $r_s^C$ and improves the Wigner-like 
character of one of the limits. On the other hand, the exponent 
$\alpha$ characterizing the average gap smoothly decays from $2$ to $1$ 
in this range of $r_s$ values. The average gap is substantially reduced 
from its HF value by the residual interaction. We associate this transition 
to a crossover in the hopping resistivity inside an insulating phase.

Partial support from the TMR network ``Phase Coherent Dynamics of 
Hybrid Nanostructures'' of the European Union is gratefully acknowledged.

\end{multicols}

\begin{references}
\bibitem{note} When $L=8$, the accuracy of our results is sufficient 
for calculating the persistent current up to $r_s = 7$. Considering 
a $2d$ torus enclosing a flux $\phi$ along the longitudinal direction, 
one finds that the transverse current (not enclosing $\phi$) is 
suppressed at $r_s^F \approx 4.8$. The $r_s^F$ obtained with $L=8$ 
and $n_e=1/16$ are very close to the one obtained for $L=6$ and 
$n_e=1/9$ (Ref. \cite{paper1}).
\end{references}
\end{document}